\newcommand{\be}{\begin{equation}}
\newcommand{\ee}{\end{equation}}
\newcommand{\bea}{\begin{eqnarray}}
\newcommand{\eea}{\end{eqnarray}}
\newcommand{\ba}[1]{\begin{array}{#1}}
\newcommand{\ea}{\end{array}}

\RequirePackage{lineno}
\documentclass[prd,aps,floats,nofootinbib,tightenlines,showpacs,linenumbers]{revtex4}
\usepackage{epsfig,graphicx,pstricks}
\usepackage{psfrag}
\usepackage{color}
\usepackage{amsmath}
\usepackage{amsfonts}
\usepackage{amssymb}
\usepackage{textcomp}
\usepackage{multirow}
\usepackage{natbib}
\usepackage{subfigure}

\begin{document}
\title {Net-Strangeness Fluctuations and Their Experimental Implications in the SU(3) PNJL Model Using the Subensemble Acceptance Method for the search of QCD Critical Point}
\author{A.~Sarkar}
\email{amal.sarkar@cern.ch}
\author{P.~Deb}
\email{paramita.dab83@gmail.com}
\author{R.~Varma}
\email{raghava.varma@cern.ch}
\affiliation{{*}School of Physical Science, Indian Institute of Technology Mandi, Kamand, Mandi - 175005, India}
\affiliation{{\dag\ddag}Department of Physics, Indian Institute of Technology Bombay, Powai, Mumbai- 400076, India}

\def\be{\begin{equation}}
\def\ee{\end{equation}}
\def\bearr{\begin{eqnarray}}
\def\eearr{\end{eqnarray}}
\def\zbf#1{{\bf {#1}}}
\def\bfm#1{\mbox{\boldmath $#1$}}
\def\hf{\frac{1}{2}}
\def\sl{\hspace{-0.15cm}/}
\def\omit#1{_{\!\rlap{$\scriptscriptstyle \backslash$}
{\scriptscriptstyle #1}}}
\def\vec#1{\mathchoice
        {\mbox{\boldmath $#1$}}
        {\mbox{\boldmath $#1$}}
        {\mbox{\boldmath $\scriptstyle #1$}}
        {\mbox{\boldmath $\scriptscriptstyle #1$}}
}
\begin{abstract}
The critical end point (CEP) is a key feature of the Quantum Chromodynamics (QCD) phase diagram, where critical phenomena cause higher-order moments of conserved charges net-baryon ($\Delta B$), net-charge ($\Delta Q$), and net-strangeness ($\Delta S$) to exhibit non-monotonic behavior. These moments and their volume-independent products are sensitive to the correlation length, making them crucial observables in the search for the CEP. In this study, we investigate net-strangeness fluctuations using the finite-volume Polyakov-loop extended Nambu–Jona-Lasinio (PNJL) model, incorporating six-quark and eight-quark interactions at energies similar to those of the RHIC beam energy scan. Our results are compared to STAR net-kaon data and the Hadron Resonance Gas (HRG) model to assess the CEP’s existence. Since direct measurement of conserved charges is experimentally challenging, net-proton, net-pion, and net-kaon are used as proxies for $\Delta B$, $\Delta Q$, and $\Delta S$. We employ the Subensemble Acceptance Method (SAM) to analyze the acceptance dependence of $\kappa\sigma^{2}$ for net-strangeness fluctuations. Our findings establish a direct mapping between the subvolume (particle fraction) and the total volume (conserved quantities), providing insights into the role of experimental acceptance in fluctuation measurements.

\end{abstract}
\maketitle
\section{Introduction}
One of the primary goals of relativistic heavy-ion collision experiments is to map the QCD phase diagram and locate the critical end point (CEP), where the first-order phase transition to the quark-gluon plasma (QGP) becomes continuous {\cite{STAR, STAR1}}. While evidence suggests the creation of hot and dense matter in the laboratory \cite{STAR3}, neither the existence nor the precise location of the CEP has been confirmed by experiments at RHIC (BNL) and SPS (CERN). In heavy-ion collisions, QGP formation is followed by expansion and freeze-out, where all chemical and kinetic interactions cease. By varying the collision energy, $\sqrt{s}$, experiments aim to bring the freeze-out point closer to the CEP \cite{STAR2_muB}. Identifying suitable observables sensitive to critical behavior, such as fluctuations of conserved quantities, is crucial in this search. Fluctuations of net-baryon $(\Delta{B})$, net-strangeness $(\Delta{S})$, and net-charge $(\Delta{Q})$ are linked to thermodynamic susceptibilities, which diverge near the CEP. However, finite-size effects limit the correlation length ($\xi$) to approximately 1.5–3 fm \cite{rajagopal2000}. Higher-order moments of conserved charges provide a robust probe of non-Gaussian fluctuations, and their non-monotonic behavior could signal the CEP \cite{rajagopal1998}. Additionally, near the CEP, critical slowing down occurs, increasing relaxation times due to the finite lifetime of the fireball \cite{koide}. Finite-size effects also influence QGP properties, as estimated through Hanbury-Brown-Twiss (HBT) radii, suggesting freeze-out volumes between 2000 and 3000 fm³ \cite{adamova}. Various QCD-based models, including the NJL, linear sigma, and Polyakov-loop NJL (PNJL) models, address finite-volume effects \cite{elze, gasser, nambu, braun, gopie, bazavov}. Studies indicate that decreasing system volume shifts the CEP to higher chemical potential $(\mu)$ and lower temperature $(T)$ {\cite{deb1, amal1, abhijit}}.

Lattice QCD calculations and QCD-inspired models \cite{boyd, engels, fodor, fukushima, ratti, pisarski, schaefer} suggest that net conserved charges ($B$, $Q$, $S$) are related to susceptibilities, $\chi_x=\langle(\delta N_x)^2\rangle/VT$. Since skewness $(S \sim \xi^{4.5})$ and kurtosis $(\kappa \sim \xi^7)$ are sensitive to the CEP, their non-monotonic fluctuations provide a crucial signature. The thermodynamic susceptibilities, $\chi_i^{(n)}$, are related to cumulants $(C_n)$ as: $\chi_i^{(n)} = \frac{1}{VT^3}C_n$, where $V$ is the system volume and $T$ is temperature. Since susceptibilities depend on system volume, moment ratios such as:
\begin{equation}
\frac{\sigma^2}{M}=\frac{\chi_2}{\chi_1}, \quad S\sigma=\frac{\chi_3}{\chi_2}, \quad \kappa\sigma^2=\frac{\chi_4}{\chi_2}
\end{equation}
are used to define volume-independent observables \cite{ejiri}.

The Subensemble Acceptance Method (SAM) is crucial for comparing experimental data with theoretical predictions \cite{40}. In heavy-ion collisions, net-kaon fluctuations serve as a proxy for net-strangeness. SAM quantifies global conservation effects and allows a direct comparison between grand-canonical susceptibilities and experimentally measured moments. For small acceptance windows, net-baryon fluctuations follow a Skellam distribution, while larger acceptance captures baryon number conservation effects \cite{41}. The acceptance fraction, $\alpha$, for net-kaon fluctuations in a subvolume is given by:
\begin{equation}
\alpha = \frac{\langle N^{acc}k\rangle}{\langle N^{4\pi}S\rangle}
\end{equation}
where $\langle N^{acc}k\rangle$ is the mean number of detected kaons. The ratio of fourth-order to second-order moments, $\kappa\sigma^2$, for net-strangeness is expressed as:
\begin{equation}\label{eq:3}
\frac{C{4}[S{1}]}{C{2}[S_{1}]}=(1-3\alpha\beta)\frac{\chi_{4}^{S}}{\chi_{2}^{S}}-3\alpha\beta \left(\frac{\chi_{3}^{S}}{\chi_{2}^{S}}\right)^{2}
\end{equation}
where $\beta = 1 - \alpha$. This framework systematically evaluates experimental acceptance effects on higher-order fluctuations.

This study investigates net-strangeness susceptibilities $(\chi_S)$ within a finite-volume, finite-density PNJL model with six-quark and eight-quark interactions. Recent heavy-ion collision studies suggest that fireball size increases with collision energy, reaching up to 10 fm for heavy systems. We analyze three finite-volume cases ($R$ = 2, 4, and 10 fm) alongside an infinite volume system, comparing our results to RHIC BES data and the Hadron Resonance Gas (HRG) model \cite{braun2001, andronic2009a}. Additionally, we apply SAM with acceptance fractions $\alpha = 0.5$–1.0 to evaluate the variation of higher-order moments $(\kappa\sigma^2)$ for net-strangeness distributions. Finally, we analyze the energy dependence of skewness $(S)$, kurtosis $(\kappa)$, and their volume-independent moment products to explore their relevance to the CEP.

\section{The PNJL model}
In this study, the 2+1 flavor PNJL model with six quark and eight quark interactions have been considered. In the PNJL model, the gluon dynamics are described by the chiral point couplings between quarks (present in the NJL part) and a background gauge field representing the Polyakov Loop dynamics. The Polyakov line is represented as,
\begin {equation}
  L(\bar x)={\cal P} {\rm exp}[i {\int_0}^\beta
d\tau A_4{({\bar x},\tau)}]
\end {equation}
where $A_4=iA_0$ is the temporal component of Euclidean gauge field $(\bar A,A_4)$, $\beta=\frac {1}{T} $, and $\cal P$ denotes path ordering. $L(\bar x)$ transforms as a field with unitary charge  under global Z(3) symmetry. The Polyakov loop is then given by 
$\Phi = (Tr_c L)/N_c$, and its conjugate by, ${\bar \Phi} = (Tr_c L^\dagger)/N_c$. The gluon dynamics can be described as an effective theory of the Polyakov loops. Consequently,
the Polyakov loop potential can be expressed as,
\begin{equation}
\frac {{\cal {U^\prime}}(\Phi[A],\bar \Phi[A],T)} {T^4}= 
\frac  {{\cal U}(\Phi[A],\bar \Phi[A],T)}{ {T^4}}-
                                     \kappa \ln(J[\Phi,{\bar \Phi}])
\label {uprime}
\end{equation}

Here, $\cal {U(\phi)}$ is a Landau-Ginzburg type potential commensurate with the Z(3) global symmetry represented by 
\cite{ratti},
\begin{equation}
\frac  {{\cal U}(\Phi, \bar \Phi, T)}{  {T^4}}=-\frac {{b_2}(T)}{ 2}
                 {\bar \Phi}\Phi-\frac {b_3}{ 6}(\Phi^3 + \bar \Phi^3)
                 +\frac {b_4}{  4}{(\bar\Phi \Phi)}^2,
\end{equation}
where, ${b_2}(T)=a_0+{a_1}exp(-a2{\frac {T}{T_0}}){\frac {T_0}{T}}$, $b_3$ and $b_4$ being constants. The second term in eqn.(\ref {uprime}) is the Vandermonde term which replicates the effect of Haar measure of the SU($N_c$) group and is given by,
\begin {equation}
J[\Phi, {\bar \Phi}]=(27/24{\pi^2})\left[1-6\Phi {\bar \Phi}+\nonumber\\
4(\Phi^3+{\bar \Phi}^3)-3{(\Phi {\bar \Phi})}^2\right]
\end{equation}
The corresponding parameters were obtained by fitting a few
physical quantities as a function of temperature in LQCD
computations \cite{ratti}. The set of values chosen here are listed in the table \ref{table1} \cite{saha}.
\begin{table}[htb!]
\begin{center}
\begin{tabular}{|c|c|c|c|c|c|c|c|c|c|c|c|}
\hline
Interaction & $ T_0 (MeV) $ & $ a_0 $ & $ a_1 $ & $ a_2 $ & $ b_3 $ &$
b_4$ & $  \kappa $ \\ 
\hline
6-quark &$ 175 $&$ 6.75 $&$ -9.0 $&$ 0.25 $&$ 0.805 $&$7.555 $&$ 0.1 $ \\
\hline
8-quark & $ 175 $&$ 6.75 $&$ -9.8 $&$ 0.26 $&$0.805$&$ 7.555 $&$ 0.1 $\\
\hline
\end{tabular}
\caption{Parameters for the Polyakov loop potential of the model.}  
\label{table1}
\end{center}
\end{table}
For the quarks, we shall use the usual form of the NJL model except for the substitution of a covariant derivative containing a background temporal gauge field. Thus the 2+1 flavor of the Lagrangian may be written as,
\begin{equation}
\begin{split}
   {\cal L} = {\sum_{f=u,d,s}}{\bar\psi_f}\gamma_\mu iD^\mu
             {\psi_f}&-\sum_f m_{f}{\bar\psi_f}{\psi_f}
              +\sum_f \mu_f \gamma_0{\bar \psi_f}{\psi_f}
       +{{g_S}\over{2}} {\sum_{a=0,\ldots,8}}[({\bar\psi} \lambda^a
        {\psi})^2+
            ({\bar\psi} i\gamma_5\lambda^a {\psi})^2] \nonumber\\
       &-{g_D} [det{\bar\psi_f}{P_L}{\psi_{f^\prime}}+det{\bar\psi_f}
            {P_R}{\psi_{f^\prime}}]\nonumber
 +8{g_1}[({\bar\psi_i}{P_R}{\psi_m})({\bar\psi_m}{P_L}{\psi_i}]^2\\
   & +16{g_2}[({\bar\psi_i}{P_R}{\psi_m})({\bar\psi_m}{P_L}{\psi_j})
          ({\bar\psi_j}{P_R}{\psi_k})({\bar\psi_k}{P_L}{\psi_i})]\nonumber
               -{\cal {U^\prime}}(\Phi[A],\bar \Phi[A],T)
\end{split}
\end{equation}

where $f$ denotes the flavors $u$, $d$ or $s$.
The matrices $P_{L,R}=(1\mp \gamma_5)/2$ are  the
left-handed and right-handed chiral projectors, respectively, and the other terms
have their usual meaning
Refs.~\cite{ghosh,deb,deb1}. The NJL part of the theory
is analogous to the BCS theory of superconductors, where the pairing of two electrons leads to condensation thereby causing a gap in the energy spectrum. Similarly to the chiral limit, the NJL model exhibits dynamical breaking of ${SU(N_f)}_L \times {SU(N_f)_R}$ symmetry to $SU(N_f)_V$ symmetry ($N_f$ being the number of flavors). As a result the composite operators ${\bar \psi_f}\psi_f$ generate nonzero vacuum expectation values (the quark condensate). The quark condensate is given as,
\begin {equation}
 \langle{\bar \psi_f}{\psi_f}\rangle= 
-i{N_c}{{{\cal L}t}_{y\rightarrow x^+}}(tr {S_f}(x-y)),
\end {equation}

where the trace is over color and spin states. The self-consistent gap equation for the constituent quark masses are,
\begin {equation}
  M_f =m_f-g_S \sigma_f+g_D \sigma_{f+1}\sigma_{f+2}-2g_1 
     \sigma_f{(\sigma_u^2+\sigma_d^2+\sigma_s^2)}-4g_2 \sigma_f^3  
\end {equation}

where $\sigma_f=\langle{\bar \psi_f} \psi_f\rangle$ denotes chiral condensate of the quark with flavor $f$. Herein,  if $\sigma_f=\sigma_u$, then $\sigma_{f+1}=\sigma_d$ and $\sigma_{f+2}=\sigma_s$, are considered, the 
expression for $\sigma_f$ at zero temperature ($T=0$) and chemical potential ($\mu_f=0$) may be written as \cite{deb},
\begin {equation}
 \sigma_f=-\frac {3{M_f}}{ {\pi}^2} {{\int}^\Lambda}\frac {p^2}{
           \sqrt {p^2+{M_f}^2}}dp,
\end {equation}

$\Lambda$ is the three-momentum cut-off value. The cut-off has been used to regulate the model because it contains couplings with finite dimensions which leads to the model being non-renormalizable. The $\Phi$, $\bar \Phi$, and $\sigma_f$ fields were obtained by the mean field approximation (MFA). However, the dynamical breaking of chiral symmetry, $N_f^2 - 1$ generates the Goldstone bosons. These Goldstone bosons are the pions and kaons whose masses, and decay widths obtained from experimental observations are further used to determine the NJL model parameters. The values of the parameters have been listed in table \ref{table2}. In the present work,  two sets of parameters for the quark sector of the PNJL model (one with six-quark interactions where $g1 = g2 =0$ and another with up to eight-quark interactions where $g1$ and $g2$ have finite values) have been considered. Moreover, it has been shown that the lowest four-quark interaction term in the quark sector forms a stable vacuum by breaking the chiral symmetry spontaneously \cite{deb}. However, the next term in the hierarchy (the six-quark interaction term), needed to mimic the $U_A(1)$ anomaly de-stabilizes the vacuum \cite{hooft,kobayashi}. Nonetheless, further increasing the interaction to eight quarks stabilizes the vacuum \cite{osipov}. Considering the stability of the PNJL model and to determine if the fluctuations have any substantial difference, it is necessary to choose two different parameter sets as elucidated in table \ref{table2}.
\begin{table}[htb]
\begin{center}
\begin{tabular}{|c|c|c|c|c|c|c|c|c|c|c|}
\hline
Model & $ m_u (MeV) $ & $ m_s (MeV)$ & $ \Lambda (MeV) $ & $ g_S \Lambda^2 $ & $ g_D \Lambda^5 $ 
&$
g_1 \times 10^{-21} (MeV^{-8})$ & $ g_2 \times 10^{-22} (MeV^{-8})$ \\ 
\hline
With 6-quark &$ 5.5 $&$ 134.76 $&$ 631 $&$ 3.67 $&$ 74.636 $&$0.0 $&$ 0.0 $ \\
\hline
With 8-quark & $ 5.5 $&$ 183.468 $&$ 637.720 $&$ 2.914 $&$ 75.968
$&$ 2.193 $&$ -5.890 $ \\
\hline
\end{tabular}
\caption{Parameters of the Fermionic part of the model.}  
\label{table2}
\end{center}
\end{table}
Once the  PNJL model has been described for infinite volume, the implementation for the finite volume constraints can be executed upon proper selection of the boundary conditions (periodic for bosons and anti-periodic for fermions). Moreover, it would lead to an infinite sum over discrete momentum values $p_i=\pi n_i/R$ ($i=x,y,z$ and $n_i$ are all positive integers and $R$ is the lateral size of the finite volume system), implying lower momentum cut-off $p_{min}=\pi/R=\lambda$.  Considering the intractable nature, the following simplifications have been made:

\begin{itemize}
\item
Surface and curvature effects have been neglected
\item
The infinite sum has been considered as an integration over a continuous
variation of momentum albeit with a lower cut-off
\item
Any modification to the mean-field parameters due to the
finite-size effects has been neglected implying a constant Polyakov loop potential as well as the mean-field part of the NJL model. Our philosophy had been to hold the known physics at zero $T$, zero $\mu$, and infinite $V$ fixed. That means we treat $V$ as a thermodynamic variable on the same footing as $T$ and $\mu$. Therefore, any variation due to the change in either of these thermodynamic parameters was translated in to the changes in the effective fields of $\sigma_f$ and $\Phi$ and through them to the meson spectra. The values of meson masses and decay constants used to fix the model parameters were thus expected to be the values strictly at $T=0$ and $\mu = 0$ and $V = \infty$. Thus the Polyakov loop potential as well as the mean field part of the NJL model would remain unchanged. They shall feel the effect of changing volume only implicitly through the saddle point equations.
\end{itemize}


The thermodynamic potential for the multifermion interactions in the mean field approximation of the PNJL model can be written as follows:
\begin {eqnarray}
 \Omega &=& {\cal {U^\prime}}[\Phi,\bar \Phi,T]+2{g_S}{\sum_{f=u,d,s}}
            {{\sigma_f}^2}-{{g_D} \over 2}{\sigma_u}
          {\sigma_d}{\sigma_s}+3{{g_1}\over 2}({{\sigma_f}^2})^2
           +3{g_2}{{\sigma_f}^4}-6{\sum_f}{\int_{\lambda}^{\Lambda}}
     {{d^3p}\over{(2\pi)}^3} E_{pf}\Theta {(\Lambda-{ |\vec p|})}\nonumber \\
       &-&2{\sum_f}T{\int_\lambda^\infty}{{d^3p}\over{(2\pi)}^3}
       [\ln\left[1+3(\Phi+{\bar \Phi}e^{-{(E_{pf}-\mu)\over T}})
       e^{-{(E_{pf}-\mu)\over T}}+e^{-3{(E_{pf}-\mu)\over T}}\right]\nonumber\\
       & +& \ln\left[1+3({\bar \Phi}+{ \Phi}e^{-{(E_{pf}+\mu)\over T}})
            e^{-{(E_{pf}+\mu)\over T}}+e^{-3{(E_{pf}+\mu)\over T}}\right]]
\end {eqnarray}

where $E_{pf}=\sqrt {p^2+M^2_f}$ is the single quasi-particle energy,
$\sigma_f^2=(\sigma_u^2+\sigma_d^2+\sigma_s^2)$ and 
$\sigma_f^4=(\sigma_u^4+\sigma_d^4+\sigma_s^4)$.

In the above integrals, the vacuum integral
has an upper cutoff of $\Lambda$, whereas the medium-dependent integrals have been extended to infinity. The lower limit of both the vacuum integral and medium-dependent integrals has been restricted down to the lower cut-off $\lambda$, where $\lambda= \pi/R $.  In view of the  6-quark and 8-quark interactions with infinite volume system and two sets of finite volume systems ($R=2 fm$ and
$R=4 fm$) each, six sets of parameter (a) PNJL-6-quark for $R=2 fm$, (b) PNJL-6-quark for $R=4 fm$, (c) PNJL-8-quark for $R=2 fm$, (d) PNJL-8-quark for $R=4fm$, (e) PNJL-6-quark for infinite volume and (f) PNJL-8-quark for infinite volume system evolved. It may be remarked here that in a strongly interacting medium, the correlation length of the colliding system (fireball) ranges between $2-3 fm$.  Therefore, the study of the observable must remain within the specified range. Nonetheless, the correlation length above the system size needs to be considered to gauge the difference in magnitude and the nature of the fluctuations both qualitatively and quantitatively. The critical temperature at $\mu=0$ MeV in 8q-PNJL model with finite volume with $R=4 fm$ is $T_c=167.0 MeV$ and for $R=2 fm$ is 
$T_c=158.0 MeV$ and in 6q-PNJL model with $R=4 fm$ is $T_c = 174.0 MeV$ and with $R=2 fm$ is $T_c = 160 MeV$. For infinite volume system the critical temperature for the 6q-PNJL model is $T_c = 181.0 MeV$ and for the 8q-PNJL model is $T_c = 168.5 MeV$.
\vspace{-0.5cm}
{\subsection{Taylor expansion of the pressure}}
The freeze-out curve $T(\mu_B)$ in the $T-\mu_B$ plane and the dependence of the baryon chemical potential on the center of mass energy in nucleus-nucleus collisions can be parametrized by \cite{cleymans}
\begin {equation}
T(\mu_B) = a - b\mu_B^2 - c\mu_B^4
\end {equation}

where $a = (0.166 \pm 0.002) $ $GeV$, $b = (0.139 \pm 0.016) $ ${ GeV^{-1}}$,  
$c = (0.053 \pm 0.021) $ $GeV^{-3} $ and
\begin {equation}
\mu_B (\sqrt s_{NN}) = d/{(1+ e\sqrt s_{NN})}
\end {equation}

with $d=1.308\pm 0.028\:GeV$, $e=0.273\pm 0.008\:GeV^{-1}$ \cite{karsch-strange}.
Conclusively, the ratio of baryon to strangeness chemical potential on the freeze-out curve showed a weak dependence on the collision energy.
\begin{equation}
{\mu_S\over\mu_B} \sim 0.164 + 0.018 \sqrt s_{NN}
\end{equation}

Although, several methods for new parametrization 
of the freeze-out curve in the $T$ vs $\mu_B$ plane have been proposed by \textit{e.g.} Borsanyi et al. \cite{borsanyi1}, the old method where the values of baryon, charge, and strangeness chemical potentials are available with respect to freeze-out temperature and BES energy has been used \cite{karsch-strange}, so as not to be restricted to the $T-\mu_B$ plane alone. The old method ensures the evaluation of all the values of the chemical potential. The pressure of the strongly interacting matter can be written as,
\begin {equation}
P(T,\mu_B,\mu_Q,\mu_S)=-\Omega (T,\mu_B,\mu_Q,\mu_S),
\label{pres}
\end {equation}

Where, $T$ is the temperature, $\mu_B$ is the baryon (B) chemical potential, $\mu_Q$ is the charge (Q) chemical potential, and $\mu_S$ is the strangeness (S) chemical potential. 
From the usual thermodynamic relations, the first derivative of pressure with respect to quark chemical potential $\mu_q$ is the quark number density. Meanwhile, the second derivative corresponds to the quark number susceptibility (QNS). Minimizing the thermodynamic potential numerically with respect to the fields $\sigma_u$, $\sigma_d$, $\sigma_s$, $\Phi$, and $\bar \Phi$, the mean field value for pressure can be obtained (\ref{pres}) \cite {deb}. The scaled pressure thus obtained in a given range of chemical potential at a particular temperature can then be expressed as a Taylor series
\begin {equation}
\frac{p(T,\mu_B,\mu_Q,\mu_S)}{T^4}=\sum_{n=i+j+k}c_{i,j,k}^{B,Q,S}(T) 
           (\frac{\mu_B}{T})^i (\frac{\mu_Q}{T})^j (\frac{\mu_S}{ T})^k
\end{equation}
\begin{equation}
c_{i,j,k}^{B,Q,S}(T)={\frac{1}{i! j! k!} 
\frac{\partial^i}{\partial (\frac{\mu_B}{T})^i} 
\frac{\partial^j}{\partial (\frac{\mu_Q}{T})^j} 
\frac{\partial^k {(P/T^4)}}{\partial (\frac{\mu_S}{T})^k}}\Big|_{\Delta \mu_{X}} 
\end{equation}

where $\Delta \mu = \mu_X - \mu_0$ and $X = B$ or $Q$ or $S$. The value of
$\mu_X$ can be obtained from the freeze-out curve. 
Further, $\mu_B$, $\mu_Q$, $\mu_S$ are related to the flavor chemical potentials $\mu_u$, $\mu_d$, $\mu_s$ as,  
\begin {equation}
  \mu_u=\frac{1}{3}\mu_B+\frac{2}{3}\mu_Q,~~~ 
  \mu_d=\frac{1}{3}\mu_B-\frac{1}{3}\mu_Q,~~~
  \mu_s=\frac{1}{3}\mu_B-\frac{1}{3}\mu_Q-\mu_S
\label{mureln1}
\end {equation}
 To extract the coefficients for baryon, charge, or strangeness fluctuations, first, the pressure is obtained as a function of $\mu_B$, $\mu_Q$ or $\mu_S$ for each value of temperature and subsequently fitted to a polynomial about $\Delta \mu_X$ using gnu plot fit program \cite{gnu}. Instead of differentiating the pressure with respect to $\Delta \mu_X$ successively to obtain the thermodynamic potential $\Omega$ (requiring higher order derivatives, up to fourth order), the Taylor expansion around $\Delta \mu_X$ has been preferred in
this study. This ensures both an easy determination and comparison of the coefficients.

The correlation coefficients have been evaluated up to fourth order,
which are generically given by;
\begin{equation}
c_{i,j}^{X,Y}=\dfrac{1}{i! j!}\dfrac{\partial^{i+j}\left(P/T^4\right)}
{{\partial\left({\frac{\mu_X}{T}}\right)^i}{\partial\left({\frac{\mu_Y}{T}}
\right)^j}}
\end{equation}

where, X and Y each stand for B, Q, and S with $X\neq Y$. For cross-correlations, the pressure can be obtained as a function of different combinations of chemical potentials for each value of temperature. The stability of the fit has been checked by varying the ranges of fit and simultaneously keeping the values of least squares to $10^{-10}$ or less.

\vskip 0.2in
\subsection{Results}
In the results, we present the higher-order cumulants (moments) of net-strangeness in the 3-flavor finite volume, finite density PNJL model. Moment products (cumulant rations) that are volume-independent are also presented.
In the Quark-Gluon Plasma (QGP), the baryon number and electric charge are carried by different quark flavors. As a result, strong correlations are expected between baryon number and charge (B-Q), charge and strangeness (Q-S), as well as baryon number and strangeness (B-S). Therefore, we also present results for the correlations among these different conserved charges. Using the 3 flavor PNJL model, data sets for various cumulants $(C_{1}, C_{2}, C_{3}, C_{4})$ have been obtained at a fixed quark chemical potential for different energies similar to the RHIC Beam Energy Scan (BES) energies (7.7, 11.5, 14.5, 19.6, 27.0, 39.0, 62.4, 130.0 and 200 GeV) to compare with experimental measured net-kaon data. In the PNJL model, three sets of finite-volume systems and the infinite-volume system have been compared with those produced in heavy-ion collisions. In the heavy-ion collision experiments, the fireball has a finite size. Hence, a finite volume system with the radius R=2fm, R=4fm, and R=10fm has been considered to enable a more accurate comparison to the experimental results.

\begin{figure*}[htb]
  \centering
   [$\mathsf{(a)}$]{{\includegraphics[width=8.2cm]{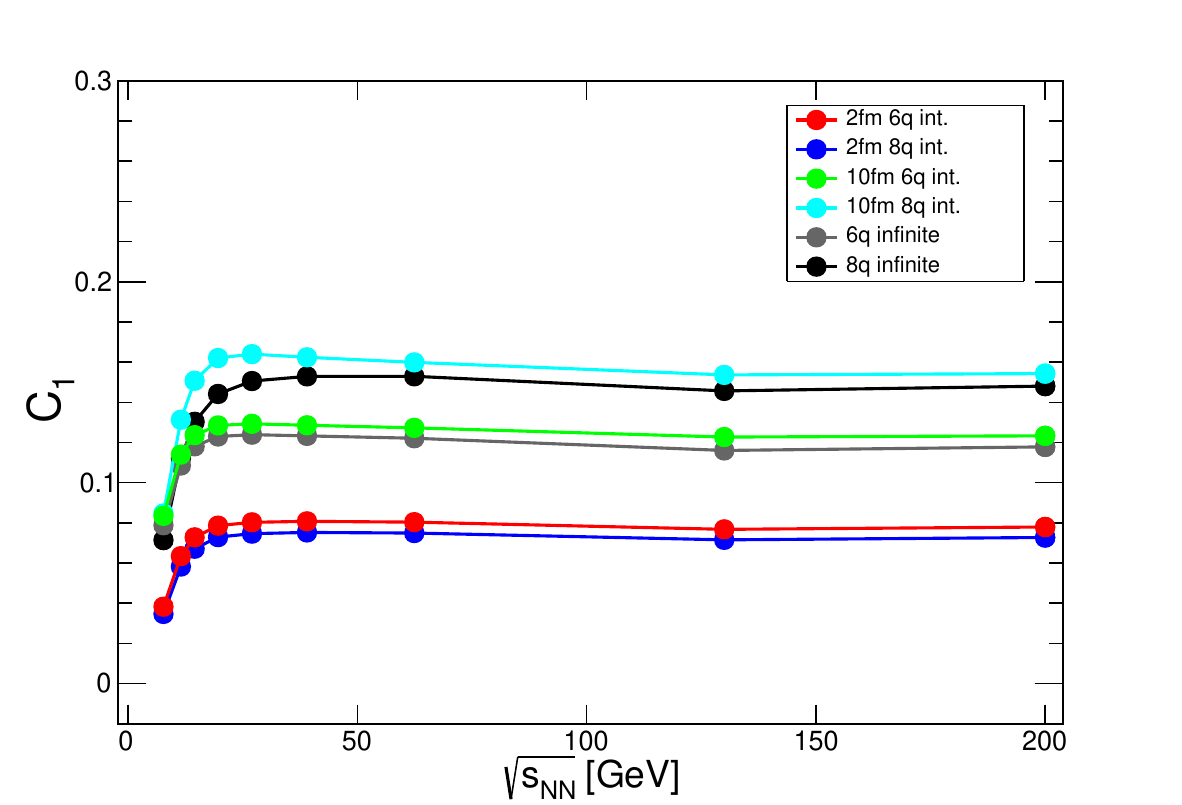} }}
   [$\mathsf{(b)}$]{{\includegraphics[width=8.2cm]{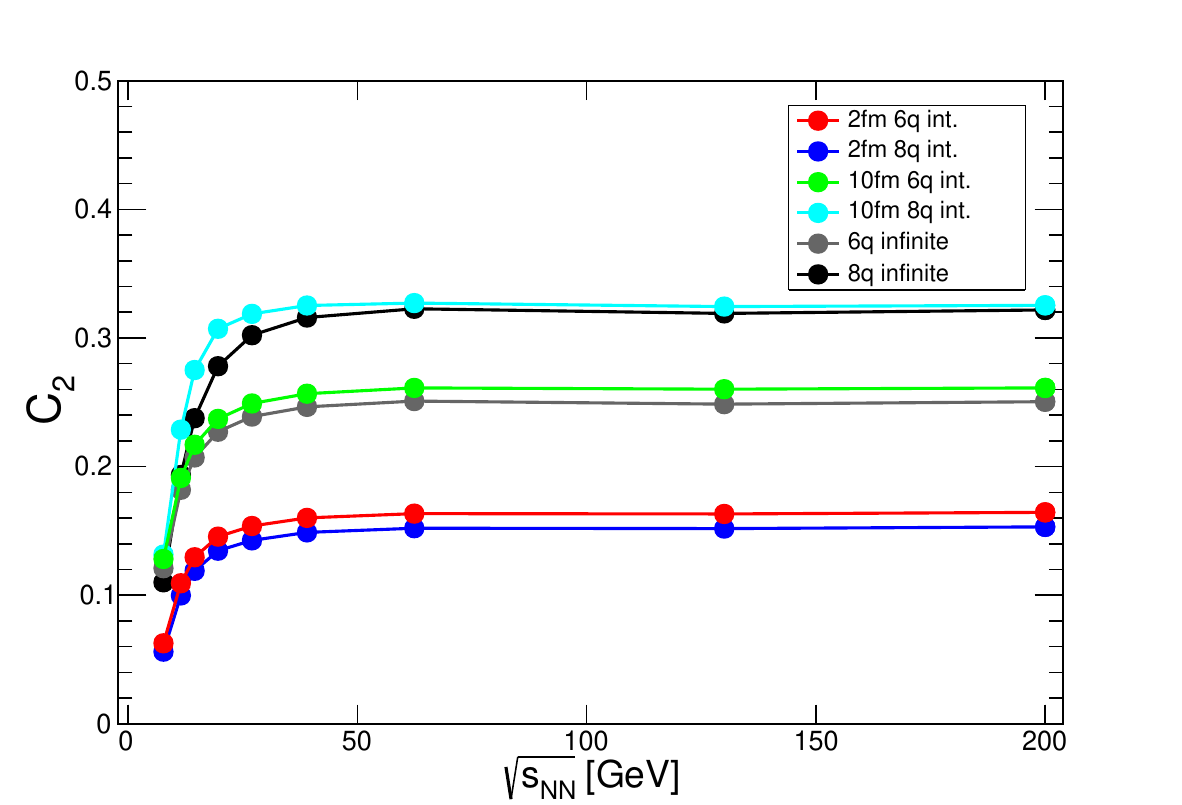} }}
   [$\mathsf{(c)}$]{{\includegraphics[width=8.2cm]{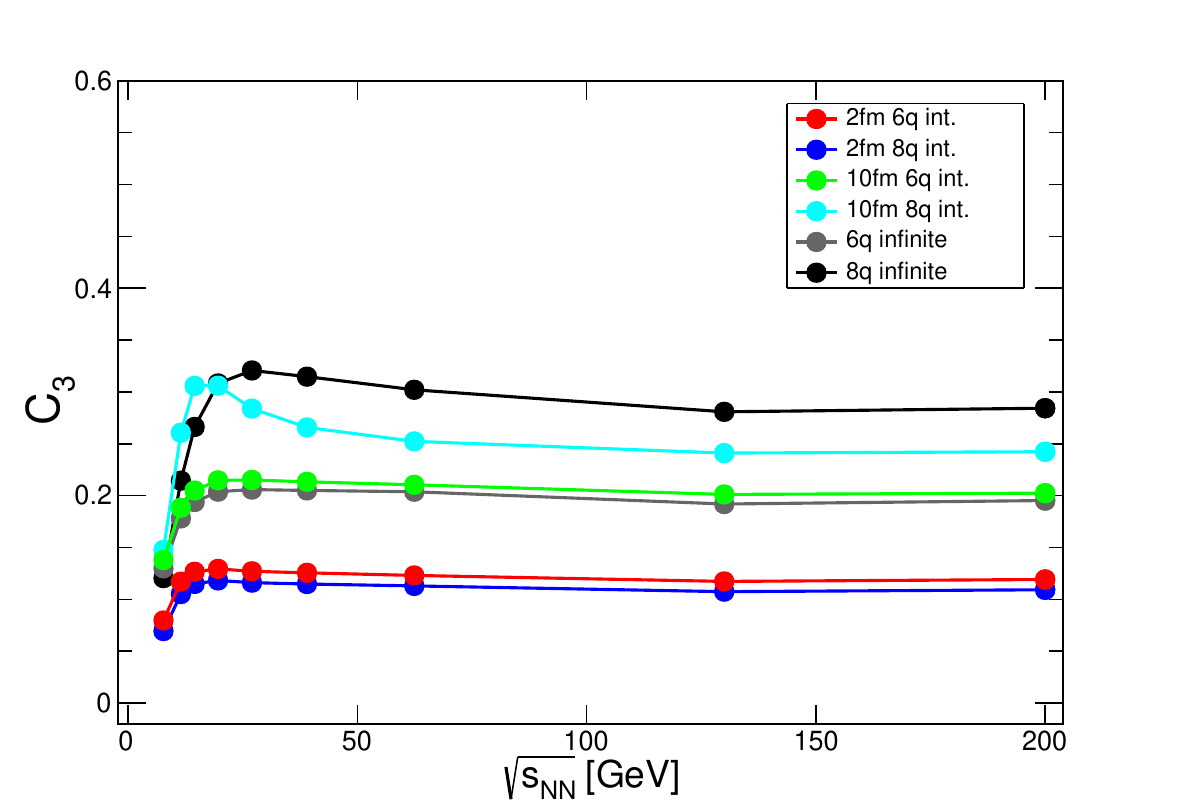} }}
   [$\mathsf{(d)}$]{{\includegraphics[width=8.2cm]{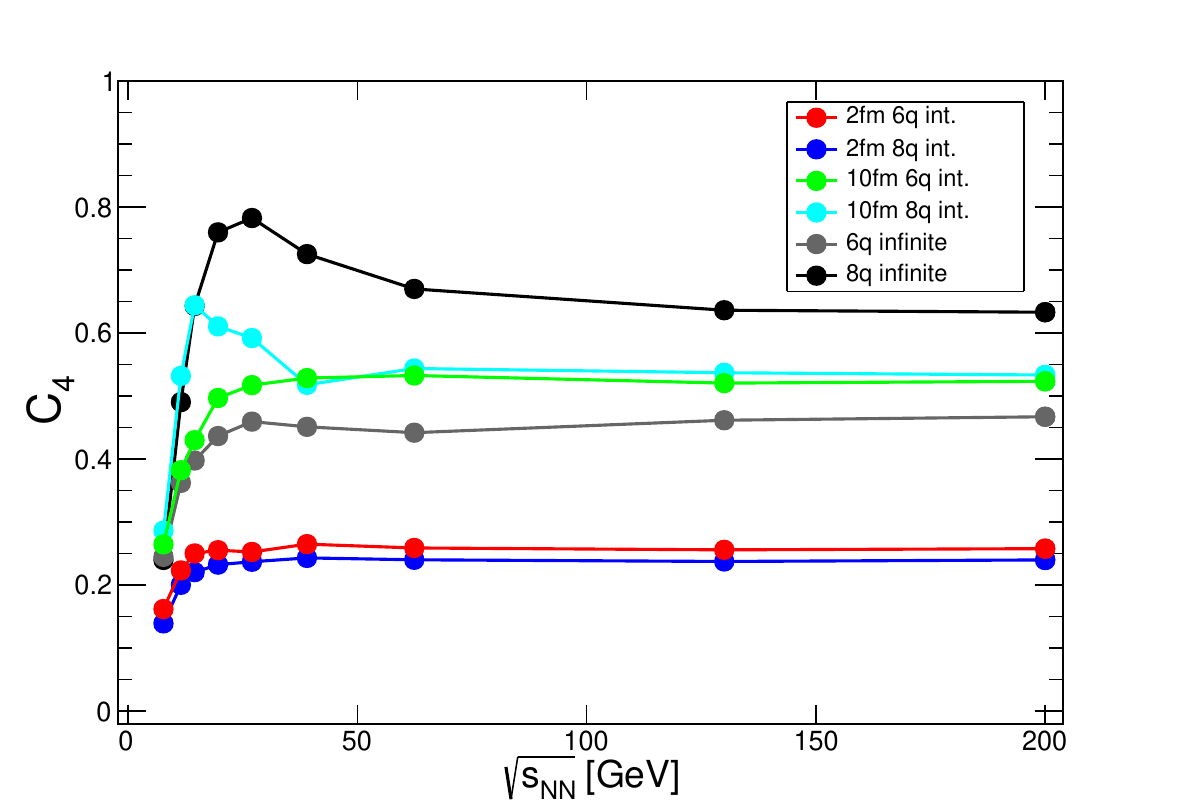} }}
    \caption{(color online) (a) First-order cumulant ($C_{1})$ (b) Second-order cumulant ($C_{2}$)  (c) Third-order cumulant ($C_{3}$) (d) Fourth-order cumulant ($C_{4}$) of net-strangeness as a function of energy in PNJL model with 6 quarks and 8 quarks interactions for infinite volume systems and finite volume systems with R = 2fm and R = 10fm. 6 quark interactions with infinite size system are shown in light grey, the finite system with R = 2fm is shown in red and R = 10fm is shown in green color. 8q interactions with the infinite size system are shown in dark grey, the finite system with R = 2fm has been shown in blue color and R = 10fm has been shown in cyan color.}
    \label{fig:1}
\end{figure*}

The Lagrangian of the Kobayashi-Maskawa–’t Hooft extended NJL model with three flavors includes interaction terms involving combinations of four-quark and six-quark interactions. Additionally, eight-quark interaction terms were later introduced to stabilize the vacuum. We are considering eight parameter sets with and without eight quark interactions (in the latter case only the six quark extension of the NJL is included) and with different volume sizes: two at 2 fm (2fm6q, 2fm8q), two at 4 fm (4fm6q, 4fm8q), two at 10 fm (10fm6q, 10fm8q) and two sets for infinite volume systems.
Various cumulants $(C_{1}, C_{2}, C_{3}, C_{4})$ calculated in the PNJL model at different energies are shown in Figure~\ref{fig:1}~[1(a),~1(b), ~1(c), and~1(d) respectively]. The system with 6 quark and 8 quark interactions with infinite size is shown in light gray and dark gray. 6q interactions for finite systems with R = 2fm and R = 10fm are shown in red and green color circles. 8q interactions for finite systems with R = 2fm have been shown in blue circles and R = 10fm have been shown in cyan circles. In Figure.~1(a), the first-order cumulants for 2fm6q and 2fm8q systems merge into a single line, increasing initially and becoming independent with increasing energy. The 10fm6q, 10fm8q finite systems, and the infinite systems show similar dependencies as in 2 fm systems with higher values in the 8 quark interaction system compared to the 6 quark interaction. The infinite systems with 6 quark and 8 quark interactions has very similar values to that of 10 fm finite systems respectively implying the system becomes very similar. Similar dependencies are observed in Figure.~1(b), for the second-order cumulants. Third-order cumulant $(C_{3})$ as a function of energy in 2fm6q 2fm8q, 10fm6q for finite volume system and 6 quark interaction for infinite system shows similar dependencies as before. Whereas, for the 10fm8q system and infinite system with 8 quark interaction $(C_{3})$ increases initially then decreases to become constant at higher energy with lower values compared to that of infinite volume. The fourth-order cumulant $(C_{4})$ as a function of energy shown in Figure.~1(d) has similar dependencies as in the third-order cumulant.

\begin{figure}[htb!]
\centering
\includegraphics[scale=0.44]{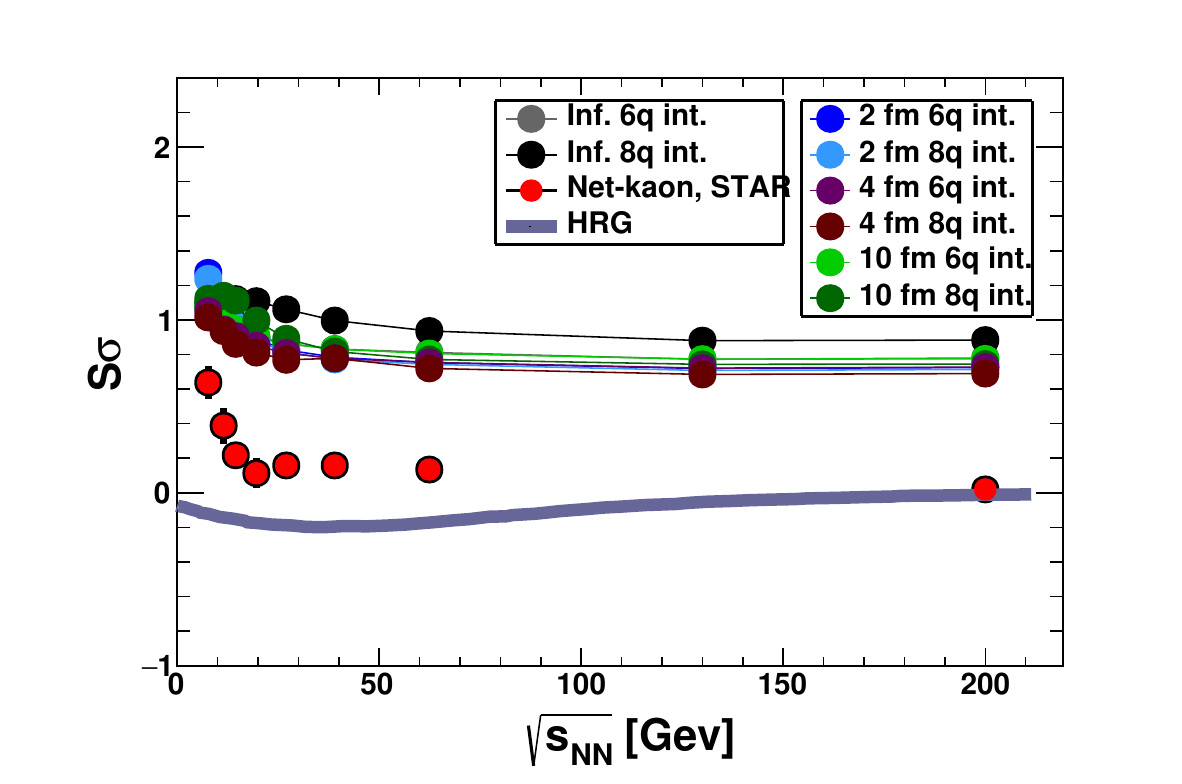}
\includegraphics[scale=0.44]{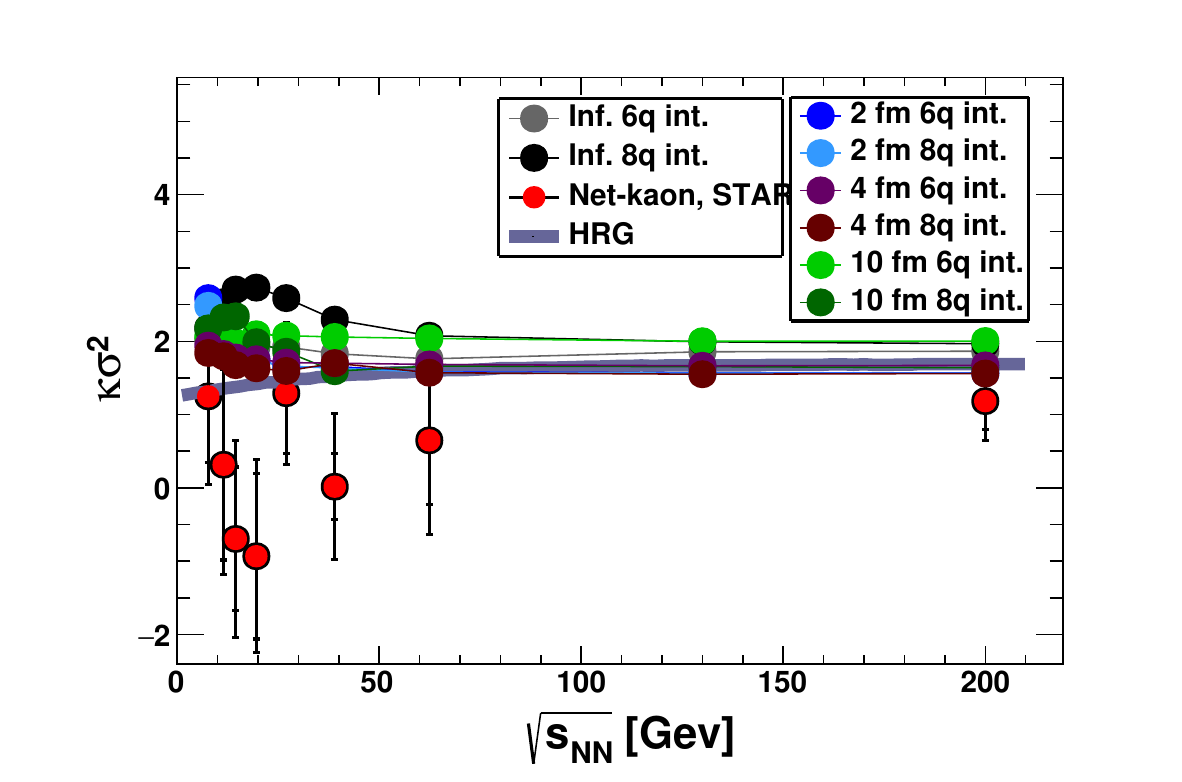}
\caption{(Color online) The moment product$ s\sigma$ ($\frac{C_{3}}{C_{2}}$) and $\kappa\sigma^{2}$ ($\frac{C_{4}}{C_{2}}$) of baryon fluctuations as a function of energy in PNJL model with 6 quarks and 8 quarks interactions for infinite volume systems and finite volume systems with R = 2fm, R = 4fm, and R = 10fm. 6 quark interactions in the system for R = 2fm are shown in blue circles, R = 4fm in purple circles, and R = 10fm in green circles. 8q interactions with R = 2fm have been shown in aqua color circles, for R = 4fm has been shown in brown color circles, and for R = 10fm in dark green color. Systems with 6 quarks and 8 quarks interaction for infinite volume are shown in gray and black color circles respectively. The results are compared with the net-kaon data of STAR experiments in red with black circles. The HRG model calculations are shown in the violate band. 
}
\label {Moment_products}
\end{figure}

Figure \ref{Moment_products} shows $s \sigma (\frac{C_3}{C_2})$ (left) and $\kappa {\sigma^2}(\frac{C_4}{C_2})$ (right) of net strangeness as a function of energy. 
Interestingly, for all combinations of size and quark interactions, both $s \sigma$ and $\kappa {\sigma^2}$ appear to be monotonic as functions of collision energy, exhibiting a local maximum for 8-quark interaction systems with 10 fm and infinite volume at the lower RHIC energies. For the moments which are volume dependent, the moment products (up to $200$ GeV) have been constructed to cancel out the dependency. Although, no large fluctuations were observed, a slightly higher value of fluctuations near lower energy (below $20$ GeV) for bigger systems was noticed. The values of $s\sigma$ and $\kappa{\sigma^2}$ are higher for infinite volume systems compare to that of finite size volumes except at the lowest RHIC energy and decreases with increasing system size. The values of $s\sigma$ are higher compared to that of STAR net-kaon results and show similar dependency as a function of energy. $\kappa{\sigma^2}$ values agreed with the experimental data at higher energies whereas fluctuations are observed at lower BES energies.

In order to make a comparison of moments calculated theoretically with the experimental data, one needs to be confident about the measured system volumes. Since measuring the system size is quite a difficult task in the experiments, one usually resorts to considering ratios of moments to eliminate the volume factor. However, this assumption is valid when interactions are small and the volume factor scales out. A recent study of the LHC energy shows that the system size of strongly interacting matter is in the range of $1.6 - 4.5 fm$. Therefore in the 
purely hadronic or partonic phases, one may observe such a scaling of the moments with system size. However, close to the critical region, such an assumption may not hold as large scale fluctuations are dominant and the system
deviates from a stable thermodynamic phase. Therefore, a comparison has also been made with the experimental data
obtained by STAR. The PNJL model data shows a similar trend as shown by the recent STAR net-kaon result. Although the model data has qualitative similarity with the experimental data, they are quantitatively different. The difference can be traced back to how both the experimental data and the model handle strangeness. While, the model calculated the net strangeness fluctuation, the experiments determined net-kaon fluctuation. Recent experimental study on net-proton fluctuation shows the possible non-monotonic behavior around $\sqrt{s} = 4$ GeV. However, the collision energy data below $\sqrt{s} =7.7$ GeV is not included in the model study due to unavailability of experimental data for net-kaon fluctuation.

\begin{figure*}[htb]
  \centering
   [$\mathsf{(a)}$]{{\includegraphics[width=8.2cm]{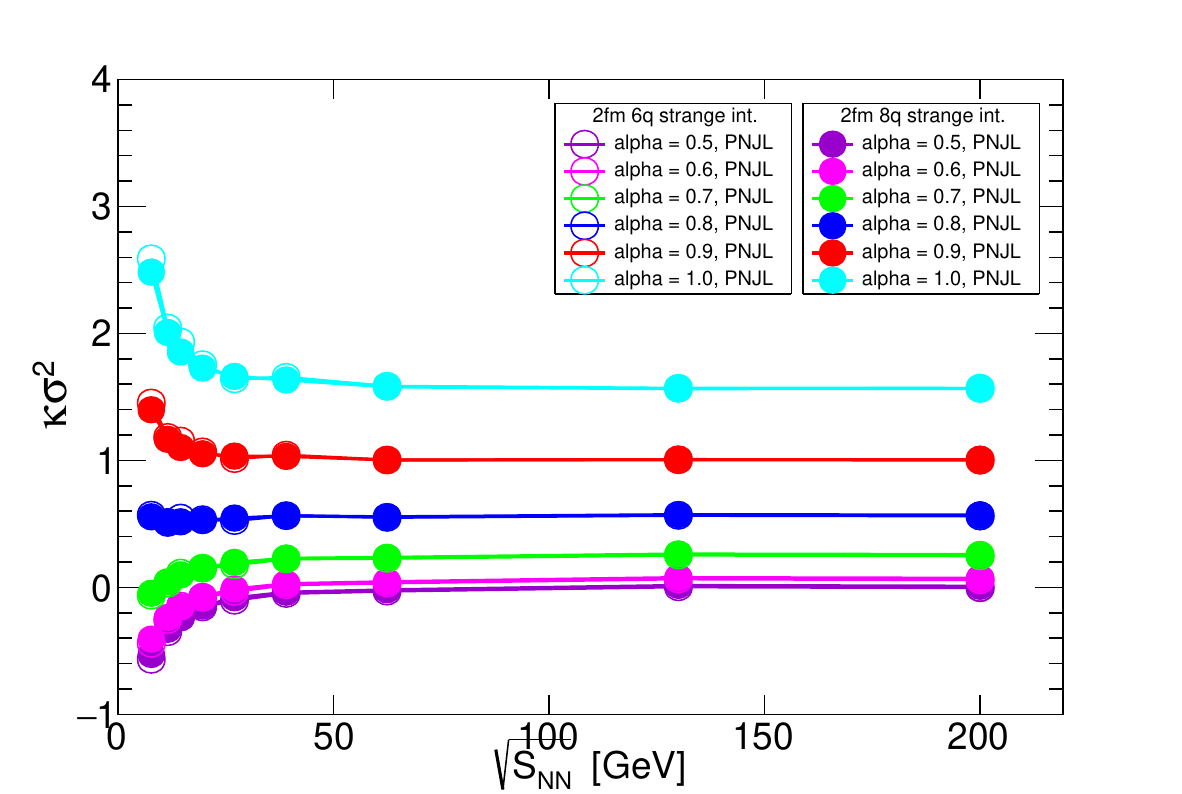} }}
   [$\mathsf{(b)}$]{{\includegraphics[width=8.2cm]{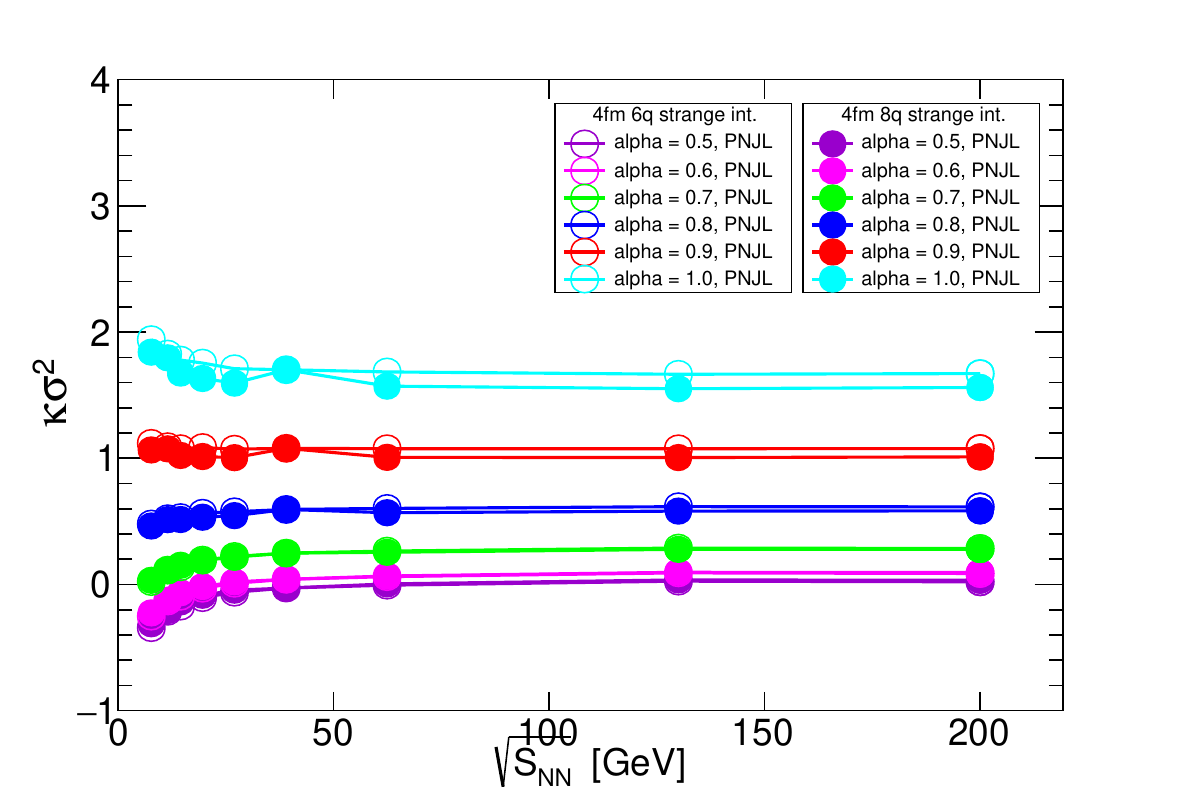} }}
   [$\mathsf{(c)}$]{{\includegraphics[width=8.2cm]{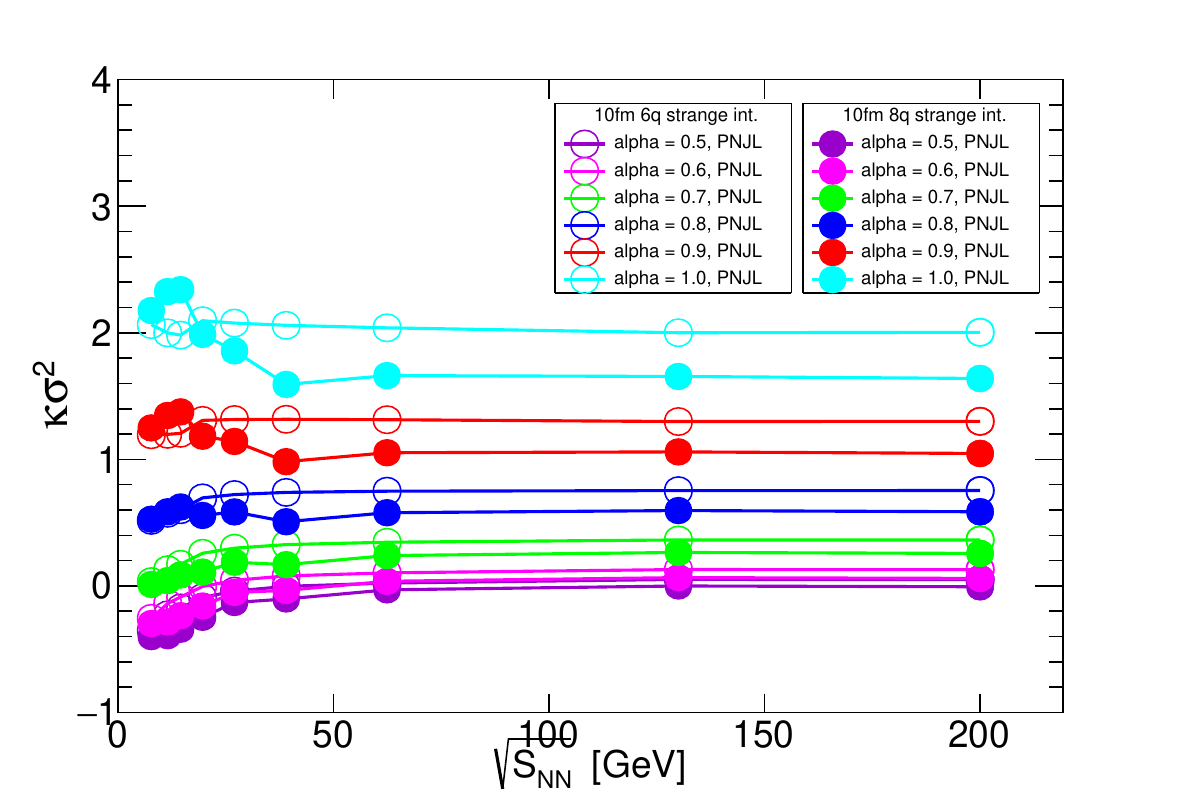} }}
   [$\mathsf{(d)}$]{{\includegraphics[width=8.2cm]{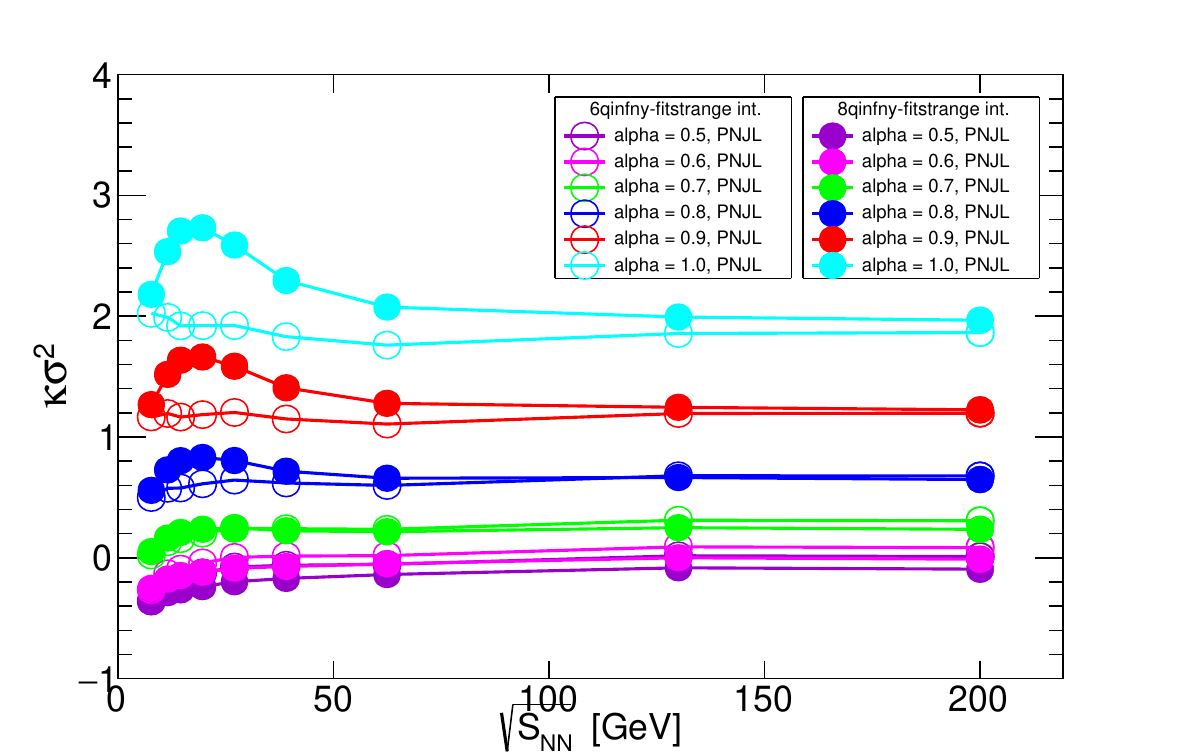} }}
    \caption{(color online) The $\kappa\sigma^{2}$ ($\frac{C_{4}}{C_{2}}$) for fixed acceptance fraction alpha $(\alpha)$ as a function of energy in PNJL model with 6 quarks and 8 quark interactions for infinite volume system and for finite volume systems with R = 2fm, R = 4fm, and R = 10fm. Solid circles represents 8 quarks interaction, and open circles represents 6 quarks interactions for different acceptance fractions}
    \label{fig:3}
\end{figure*}

Fig.~\ref{fig:3} shows $\kappa\sigma^{2}$ of net-strangeness as a function of energy for fixed acceptance fraction $(\alpha)$ for a system with 6 quark and 8 quark interactions for both finite and infinite volume. Open circles represents 6 quark interaction systems and solid circles represents 8 quark interaction system in the PNJL model. The result for $\alpha$ - dependence of $\kappa\sigma^{2}$ is calculated by using Eq. (\ref{eq:3}) of the Sub-ensemble Acceptance Method. The graph displays the value of $\kappa\sigma^{2}$ within the sub-volume system that is obtained from the total net strange number at a particular energy, with each data point representing the acceptance percent. 


At lower energy regimes, the $\alpha$ dependent results exhibit small fluctuations; however, as energies increase, the results remain energy-independent. This figure also demonstrates that the fluctuation at lower energy regions diminishes as the value of alpha ($\alpha$) decreases.

\vspace{-0.5cm}
\subsection{Phase Diagram}

The location of the critical endpoint in the phase diagram is a fundamental question in the study of hot and dense strongly interacting matter. Phase diagrams are typically determined by identifying the critical temperature ($T_c$) as the temperature at which the light quark chiral condensate exhibits a discontinuity or by locating the maximum of the derivative of the light quark condensate to temperature for various chemical potentials ($\mu$). The CEP serves as the boundary separating the crossover transition from the first-order transition.
\begin{figure}[htb]
\centering
\includegraphics[scale=1.1]{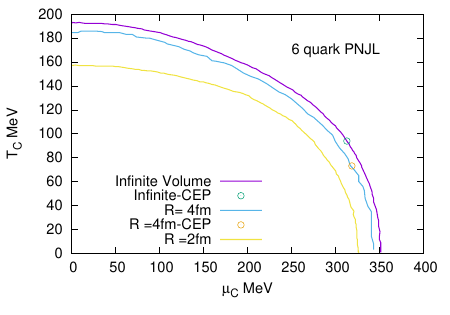}
\includegraphics[scale=1.1]{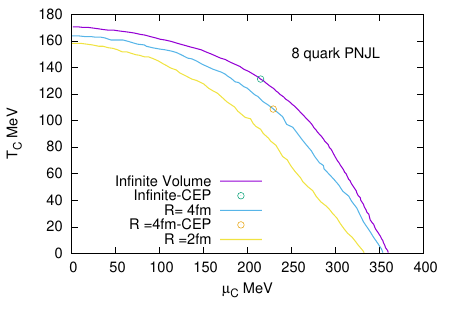}
\caption{(Color online) Phase diagram of the 6q (left) and 8q (right) PNJL model for infinite volume and finite volume system with $R=4fm$ and $R=2fm$.}
\label {tc-mub}
\end{figure}
Significant efforts in lattice QCD and QCD-inspired models have been dedicated to identifying the CEP in the $T-\mu$ phase diagram. This study presents a comparative analysis of phase diagrams for six-quark and eight-quark PNJL models in both infinite and finite volume systems. The results for six-quark and eight-quark are shown in the left and right panel of Figure~\ref{tc-mub}. In the region where $T < T_{CEP}$ and $\mu > \mu_{CEP}$, the chiral and de-confinement transitions are of first order. The CEP locations for the infinite volume system and the finite volume system with R = 4 \, \text{fm} are provided in Table~\ref{table3}.
\begin{table}[htb]
\begin{center}
\begin{tabular}{|c|c|c|c|c}
\hline
Model & $ (\mu_c, T_c) (MeV) $ \\
\hline
PNJL6q-infinite-volume &$ (312.5, 93.9) $   \\
\hline
PNJL6q-4fm & $ (318.5, 73.1) $ \\
\hline
PNJL-8q-infinite-volume & $(214.8, 131.2) $\\
\hline
PNJL-8q-4fm & $(229.2, 108.6) $\\
\hline
\end{tabular}
\caption{The values of CEP $(\mu_c, T_c)$ for different sets of PNJL model.}
\label{table3}
\end{center}
\end{table}
From the analysis, it is observed that the critical temperature decreases, and chemical potential increases with a reduction in system size. Consequently, the CEP shifts to progressively lower temperatures as the volume decreases, ultimately vanishing at R = 2 \, \text{fm}. At R = 2 \, \text{fm}, the chirally broken phase completely disappears, leading to a crossover-like transition. For an infinite volume system, the CEP shifts toward higher chemical potentials, and the first-order transition line contracts with decreasing system size until it meets the axis at R = 2 \, \text{fm}, where the system size becomes comparable to the confinement scale. 

\section{summary}
We have discussed the properties of the net strangeness fluctuations in the nuclear matter using the PNJL model where the skewness and the kurtosis are considered to be the observables for CEP. All the correlations were obtained by fitting the pressure in a Taylor series expansion around the strangeness chemical potentials. These chemical potentials are obtained from the freeze-out curve which depends on the collision energies in the BES scan at the RHIC heavy-ion collision experiment. The results are shown for the PNJL model with 6 quark and 8 quark interactions and for infinite volume and two finite volume systems with lateral size of $R=2 fm$, $R=4fm$, and $R=10fm$ are considered for this study. Interestingly, the values of the fluctuation do not show any major differences for both the 6-quark and 8-quark interactions, but they are sensitive to the system size. Furthermore, both the skewness and the kurtosis of the strangeness fluctuation in the PNJL model developed here have similar features, in line with the collision
energies reported in the heavy ion experiments. As the temperature is increased both the skewness and the kurtosis values decrease. Further, an enhancement of fluctuations for low collision energy have been found. It can be observed that the fluctuation is more for large system sizes. The volume dependence shows an expected scaling behavior in the hadronic and partonic phases. In the critical region, the system size scaling breaks down and may be used to estimate the 
closeness of the created fireball to the critical region. Generally, the leading order coefficients can be most useful in identifying the formation of QGP, while the higher order coefficients can help identify the crossover region.

Nonetheless, in this simpler version of finite volume effects the strength of the fluctuations at the lowest end of the energies may be also due to the exclusion of modes which can be improved by accounting modes with momenta below the confining scale. In the higher-order moments study of strangeness fluctuation, the correlation length of the system is comparable to the finite system size of the system, and therefore negligible fluctuation has been observed compared to the infinite volume system where the correlation length is not comparable to the system size. 
The phase diagram for the 6-quark and 8-quark PNJL model has also been discussed for infinite volume and finite volume systems with $R=2fm$ and $R=4fm$ and the critical end point for different system sizes has been discussed. As the system size decreases, the position of the critical points shifts towards the lower temperature and chemical potential until it vanishes completely at $R=2fm$.

\acknowledgements P.D would like to thank Women Scientist Scheme A (WOS-A) of the Department of Science and Technology (DST) funding with grant no SR/WOS-A/PM-10/2019 (GEN). Part of the work has been presented at the Light Cone Conference 2017 \cite{DebFBS}.  
\\
\\

\end{document}